\documentclass[english,
aps, prb, reprint, 
sort&compress, showpacs]{revtex4-1}

\usepackage{amsmath}
\usepackage{graphicx}
\usepackage[
dvipdfm,
colorlinks=true, citecolor=blue, linkcolor=blue, urlcolor=blue]
 {hyperref}

\begin{document}

\title{Control and Dynamic Competition of Bright and Dark Lasing States\\
in Active Nanoplasmonic Metamaterials}

\author{Sebastian Wuestner, Joachim M.~Hamm, Andreas Pusch, Fabian Renn, Kosmas L.~Tsakmakidis, and Ortwin Hess}

\email{o.hess@imperial.ac.uk}

\affiliation{The Blackett Laboratory, Department of Physics, Imperial College London, South Kensington Campus, London SW7 2AZ, United Kingdom}

\begin{abstract}
Active nanoplasmonic metamaterials support bright and dark modes that compete for gain. Using a Maxwell-Bloch approach incorporating Langevin noise we study the lasing dynamics in an active nano-fishnet structure. We report that lasing of the bright negative-index mode is possible if the higher-$Q$ dark mode is discriminated by gain, spatially or spectrally. The nonlinear competition during the transient phase is followed by steady-state emission where bright and dark modes can coexist. We analyze the influence of pump intensity and polarization and explore methods for mode control.
\end{abstract}


\maketitle

Plasmonic nanolasers,\citep{Hill2007,Noginov2009,Oulton2009,Berini2011} arguably the smallest attainable laser devices, are precursors towards active integrated nano-optics. Owing to their ability to bind and coherently amplify light, these subwavelength devices may also be regarded as coherent ``meta-atomic'' sources, elementary building blocks of light-emitting metamaterials. In isolation, a nanolaser utilizes feedback by stimulated emission into a strongly localized, weakly radiative resonance of a metal-dielectric nanoparticle \citep{Noginov2009} or nanostructure.\citep{Oulton2009,Ma2011} For deep-subwavelength particles radiative loss becomes negligible and the nanolaser resonates with modes that are effectively trapped on the metal surface. In this limit, the nanolaser is a spaser,\citep{Bergman2003}  a dark coherent emitter of plasmons that can be theoretically described in quasistatic approximation. Consequently, the spasing threshold is reached when gain overcomes dissipative loss and only depends on the intrinsic dispersion of the materials but not on the geometry.\citep{Wang2006,Stockman2010}

The introduction of gain into optical metamaterials, on the other hand, has been primarily motivated by the aim to design loss-free metamaterials.\citep{Xiao2010,Wuestner2010,Fang2010,Meinzer2010,Soukoulis2011} Beyond the point of loss compensation, when gain outweighs dissipative loss, the structure becomes amplifying.\citep{Hamm2011} Higher gain densities (and pump intensities) may lead into the lasing regime if dark modes are suppressed and dissipative and radiative losses are overcome.\citep{Zheludev2008,Tanaka2010} The required feedback emerges from the coupling of the bright and dark meta-atomic resonances to Bloch bands $\omega^{(i)}(k)$, which can differ greatly in terms of their damping constants. With increasing pump strength, and depending on the band dispersion and the gain spectrum $g(\omega)$, an increasing number of $k$ states becomes unstable (freely oscillating), entering the nonlinear competition for gain. If dispersion is weak (flat bands), the $k$ states will equally benefit from the gain and spatial hole burning effects will dominate, causing a complex spatio-temporal light-field dynamics known as filamentation.\citep{Hess1996} Hence, two aspects require clarification. First, more fundamentally, whether coherent light emission is possible in nanoplasmonic gain-enhanced metamaterials despite the unique presence of dark modes, which may clamp the gain before any bright mode can cross the threshold. Beyond that a second, technically relevant, aspect concerns the control of spatial coherence by engineering the dispersion of the  $\omega^{(i)}(k)$ bands to effectively suppress certain Bloch states, enabling, e.g.~narrow angle emission from large-area surfaces. 

\begin{figure}
 \begin{centering}
  \includegraphics{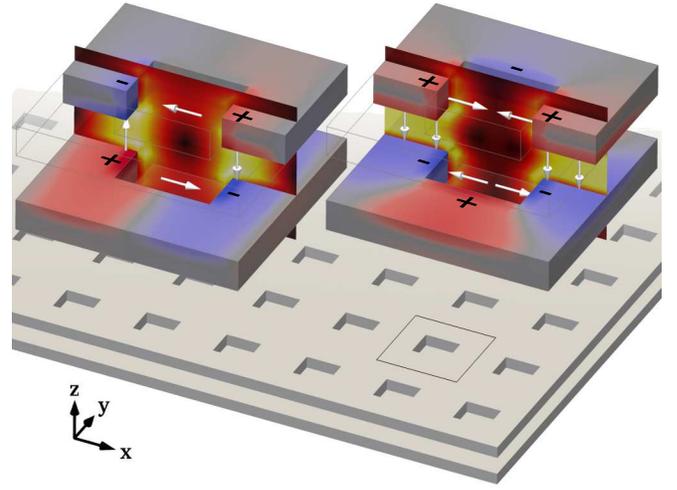}
 \end{centering}
 \caption{(color online) Electric field enhancements and charge distributions on the metal films inside the double fishnet unit cell for the bright negative-index mode (left) and the dark quadrupole mode (right). The direction of the electric field in the $x$-$z$ plane is indicated by the white arrows.\label{Figure1}}
\end{figure}

In this Rapid Communication, we investigate the fundamental nonlinear dynamic interplay of bright and dark lasing states in an active nanoplasmonic metamaterial (see Fig.~\ref{Figure1}).
To obtain insight into the nature of this competition, we reduce the potentially overwhelming complexity of the dynamic many-mode interplay to $k_{\parallel}=0$ (normal to plane) emission, where the dipolar mode is brightest and the quadrupolar mode is truly dark (does not radiate). A four-level Maxwell-Bloch approach for gain-enhanced nanoplasmonic metamaterials, previously employed to study loss-compensation in nano-fishnet metamaterials in an ultrafast pump-probe setup \citep{Wuestner2010,Wuestner2011} and in an amplification setting,\citep{Hamm2011} is extended by stochastic Langevin forces \citep{Pusch2012,supplement,Andreasen2009,Andreasen2011,Drummond1991} to accommodate for bath-induced quantum noise.\citep{Gardiner1991} This microscopic \emph{ab initio} approach takes into account all relevant scattering channels and incorporates the two prevalent forms of light and plasmon emission, amplified spontaneous emission (ASE) and coherent stimulated emission, into a self-consistent theoretical framework. 

\begin{figure}
 \begin{centering}
   \includegraphics{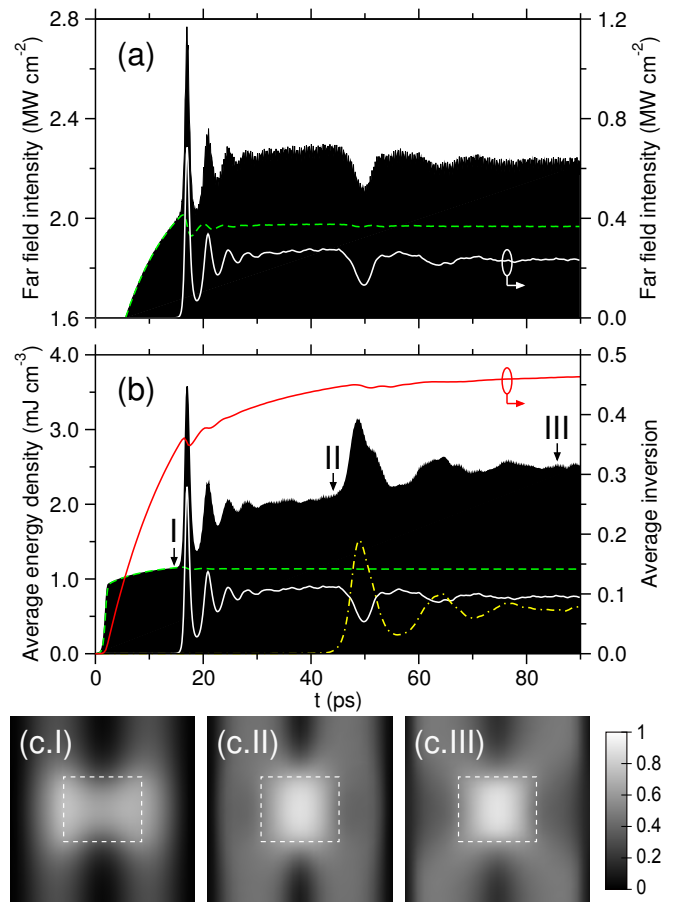}
 \end{centering}
 \caption{(color online) (a) Far-field intensity in transmission and (b) energy and average inversion (red solid line, right axis) inside the nano-fishnet metamaterial over time. The signals, time-averaged over $0.4\,\text{ps}$ (black), are decomposed into the pump mode at $680\,\text{nm}$ (green dashed line), the bright mode at $717.25\,\text{nm}$ (white solid line), and the dark mode at $731.8\,\text{nm}$ (yellow dash-dotted line). (c) Inversion profiles in the $x$-$y$ plane centered between the metal films at the times indicated in (b).\label{Figure2}}
\end{figure}

Our investigations focus on a double-fishnet structure with unit-cell side length of $p=280\,\text{nm}$ and hole side lengths of $120\,\text{nm}$ and $100\,\text{nm}$ in $x$ and $y$ directions, respectively. The thickness of the two holey silver films is $40\,\text{nm}$ and the spacer layer between the films is $60\,\textrm{nm}$ thick. Both the spacer layer and the holes are filled with a dielectric of refractive index $n=1.62$. The (cold-cavity) $k_{\parallel}=0$ modes of the passive structure are a bright mode at $713.8\,\text{nm}$ with $Q$ factor $Q=52$ and a dark mode at $732.2\,\text{nm}$ with $Q=134$. Figure \ref{Figure1} depicts the enhancement and direction (white arrows) of the electric field for both modes in the $x$-$z$ plane. A further analysis of the passive modes can be found in the Supplemental Material.\citep{supplement,harminv,*Mandelshtam1997,Mary2008,Yang2011} Whilst the two modes have different shapes, they overlap in the vicinity of the hole and will, in active structures, compete for gain. To explore the nonlinear dark-bright mode dynamics above threshold, gain is introduced into the spacer region and the holes of the structure. We describe the gain medium by a four-level system absorbing at $680\,\text{nm}$ with a coupling strength of $\sigma_{a}=1.3\,\text{C}^{2}\text{kg}^{-1}$ and emitting at $\lambda_{e}=718\,\text{nm}$ with a coupling strength of $\sigma_{e}=1.03\,\text{C}^{2}\text{kg}^{-1}$. The density of the gain molecules is set to $N=3\cdot10^{19}\,\text{cm}^{-3}$ and the spectral width of the emission line is chosen to be $\Gamma_{e}=1/\left(20\,\text{fs}\right)$; further gain parameters and a detailed description of the four-level model have been presented elsewhere.\citep{Wuestner2011,supplement} We assume that the system is continuously pumped with $E_{x}$-polarized (later $E_{y}$-polarized) light at a wavelength of $680\,\text{nm}$ and that at $t=0$ all gain molecules reside in their ground state.
In a first numerical experiment, we study the transient behavior at a pump intensity of $10.75\,\text{MWcm}^{-2}$, well above threshold. Figure \ref{Figure2} 
shows the temporal evolution over the first $90\,$ps of the total signals (black areas), the filtered signals (solid white, dashed green and dashed-dotted yellow lines) in both far-field [Fig.~\ref{Figure2}(a)] and near-field [Fig.~\ref{Figure2}(b)], as well as the average inversion [red solid line in Fig.~\ref{Figure2}(b)]. At early times ($t<17\,\text{ps}$), the total far-field intensity, composed only of the transmitted pump field, steadily rises due to saturation of the absorption. During this phase the internal inversion also continuously rises. After approximately $18\,\text{ps}$, a strong burst is observed in the far-field, initiating relaxation oscillations of the intensity and inversion on a time-scale of $3.8\,\text{ps}$. Initially, these oscillations seem to lead to stable steady-state emission but are then disrupted by large intensity dips starting at $t=45\,\text{ps}$. Following these dark oscillations, emission only steadies again at around $t=80\,\text{ps}$.

Using a time-dependent frequency filter,\citep{supplement,Chen1993,Zwillinger2003} we decompose the far-field signal into its frequency components. Two dominant frequencies are found: The pump at $680\,\text{nm}$ (green dashed line) and, subject to frequency pulling,\citep{Siegman1986} the bright mode at $717.25\,\text{nm}$ (white full line). Figure~\ref{Figure2}(a) clearly shows that the transmitted pump intensity stabilizes as the bright relaxation oscillations dampen in the far-field. The origin of the intensity oscillations at $50\,\text{ps}$ is not obvious from the far-field measurements alone, yet we observe that the bright mode settles at a lower intensity value after these dips. To understand this behavior we resolve the near-field dynamics by recording and filtering the temporal evolution of the internal energy density [Fig.~\ref{Figure2}(b)]. This reveals that the dips in the far-field are associated with relaxation oscillations of the internal dark mode at $731.8\,\text{nm}$ (yellow dash-dotted line) with a frequency of $\sim\!\!\left(15\,\text{ps}\right)^{-1}$. Strong mode competition causes the bright mode intensity to drop when the dark mode peaks during its relaxation oscillation. As the bright and dark modes undergo relaxation oscillations, the distribution of inversion changes due to spatial hole burning effects. In Fig.~\ref{Figure2}(c) we show the profiles in the $x$-$y$ plane extracted at three different points in time: (I) just before the first spike of the bright mode, (II) before the first spike of the dark mode, and (III) when the bright and dark mode energies inside the resonator have settled. Before lasing sets in, inversion is deposited in regions where the pump mode has high intensity values [Fig.~\ref{Figure2}(c.I)]. When the bright mode becomes unstable, inversion is rapidly depleted in regions where its intensity is highest, while the pump keeps providing inversion to other regions [Fig.~\ref{Figure2}(c.II)]. Eventually, with the inversion still rising, the dark mode receives enough gain to become unstable, further reducing the inversion around the hole. After this point, the highest inversion is found within a small square area around the center, where stimulated emission is weak as both modes have a nodal point in the center of the unit cell [Fig.~\ref{Figure2}(c.III)].

\begin{figure}
 \begin{centering}
   \includegraphics{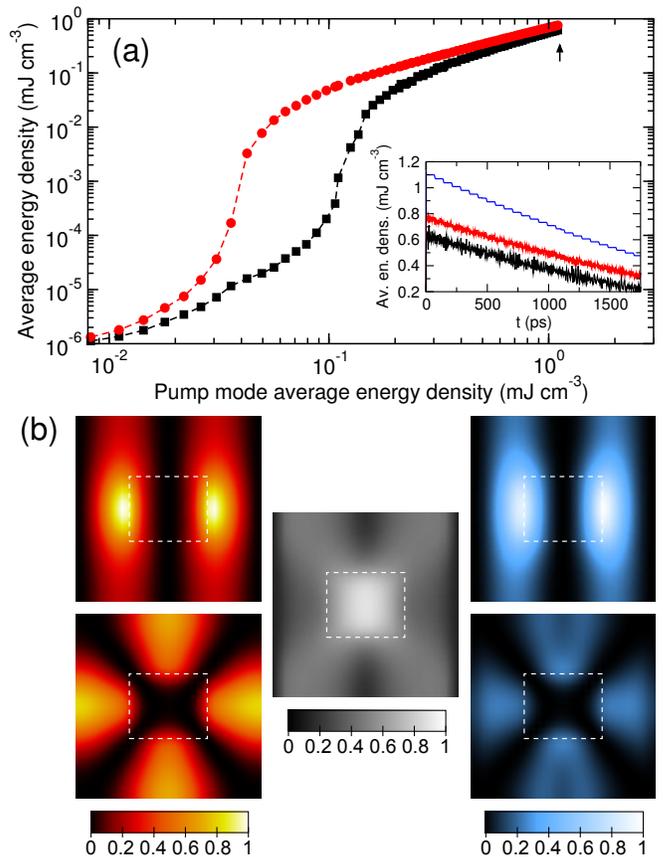}
 \end{centering}
 \caption{(color online) (a) Dependence of the average energy densities of the bright (red dots) and dark mode (black squares) on the pump mode average energy density. The inset illustrates how the energy densities follow the ramp-down of the pump starting at the pump intensity marked by the arrow. (b) Mode intensity (left), inversion (middle) and local gain [right, see Eq.~\eqref{eq:localgain} in the $x$-$y$ plane centered between the metal films.\label{Figure3}}
\end{figure}

The fact that the bright mode starts oscillating first indicates that this mode has a lower threshold than the dark mode. To further understand the threshold behavior and the relative strengths of the dark and bright modes, we numerically extract the steady-state intra-cavity energies for varying pump intensity. We first pump the system with high intensity, effectively reducing the damping time of the relaxation oscillation that is inversely proportional to the pump intensity. By ramping down the pump power with increasingly longer steps of $70-210\,\mathrm{ps}$, we ensure that the modal energies can adiabatically follow without pronounced relaxation bursts [inset in Fig.~\ref{Figure3}(a)].
The so extracted input-output curves for the two modes, log-log plotted in Fig.~\ref{Figure3}(a), show an asymptotic linear dependence on the pump power above and below threshold. Owing to the presence of amplified spontaneous emission, a gradual transition to lasing occurs for both the bright and the dark mode. Within our modelling approach spontaneous emission enters in the form of Langevin noise originating from the strong system-bath coupling in the gain medium. Indeed, our results confirm that the induced dipole fluctuations qualitatively reproduce the threshold behavior predicted by (class-B) rate equation models  that incorporate spontaneous emission in the form of phenomenological terms, which account for the coupling of inversion to the ground-state of the quantized photon field. The threshold of the bright mode is reached at a pump energy density of approximately $0.03\,\text{mJcm}^{-3}$ (corresponding to a pump intensity of $0.3\,\text{MWcm}^{-2}$). In contrast, the threshold of the dark mode is crossed at a pump energy density of around $0.1\,\text{mJcm}^{-3}$ ($1.1\,\text{MWcm}^{-2}$), more than three times higher than that of the bright mode. 

Thus, despite the higher $Q$ factor of the dark mode, it is the bright mode that reaches threshold first. This implies that, for the chosen set of optogeometric parameters, the gain provided to the bright mode is significantly higher compared to that available to the dark mode. In general, the local gain is both time- and frequency-dependent and can be defined as
\begin{equation}
 g_{loc}(\omega,\textbf{r},t)=\frac{\sigma_{e}\mathcal{L}(\omega)\left|E(\omega,\textbf{r})\right|^{2}\Delta N(\textbf{r},t)}{2\Gamma_{e}\int_{V}u(\omega,\textbf{r})\text{d}^{3}\mathbf{r}}\:,\label{eq:localgain}
\end{equation}
where $\mathcal{L}(\omega)$ is the Lorentzian emission line and $u$ is the energy density. Accordingly, the overlap of the intensity profile with the inversion strongly affects the gain delivered to the modes. In the left column of Fig.~\ref{Figure3}(b) we depict the normalized intensity profiles $\left|E(\omega,\textbf{r})\right|^{2}/\left|E_{max}\right|^{2}$ of the bright (top) and dark (bottom) modes in the $x$-$y$ plane. The plot in the middle shows the steady-state distribution of the inversion $\Delta N(\textbf{r},t\rightarrow\infty)$ in the same plane. Multiplying the two profiles with the emission line results in the local gain profile at steady state $g_{loc}(\omega,\textbf{r},t\rightarrow\infty)$ [right column in Fig.~\ref{Figure3}(b)]. In this way, it can be clearly seen that the gain rate, given by the volume integral over the local gain, is significantly higher for the bright mode, as expected.

\begin{figure}
 \begin{centering}
  \includegraphics{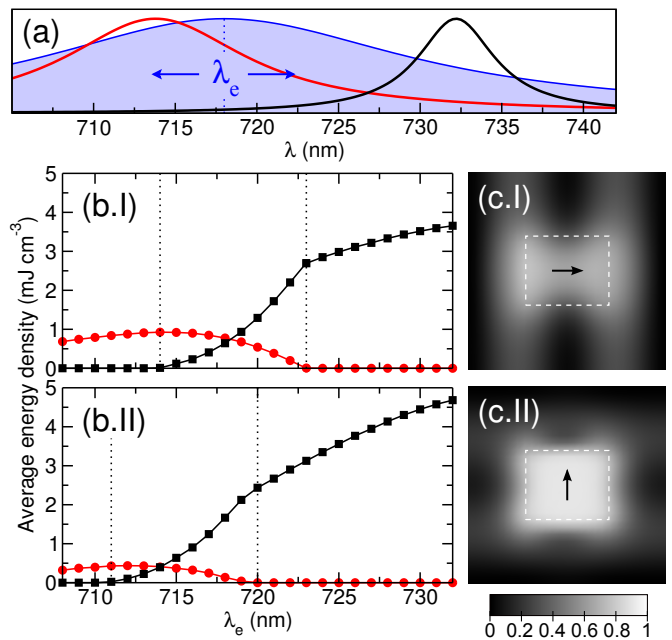}
 \end{centering}
 \caption{(color online) Dependence of steady-state emission into dark and bright modes on emission wavelength and polarization. 
(a) The emission line (here displayed for $\lambda_{e}=718\,\text{nm}$) and dark and bright mode lines are shown schematically indicating the maximum emission at $\lambda_{e}$ and how it is shifted for control. 
(b) Average mode energy densities during steady state in dependence of the emission wavelength of the dye for $E_{x}$ (I) and $E_{y}$ polarization (II) of the pump field with intensity $10.75\,\text{MWcm}^{-2}$. The bright mode is given by the red line and dots, the dark mode by the black line and squares. 
(c) Inversion profiles before the onset of relaxation oscillations for $E_{x}$ (I) and $E_{y}$ polarization (II) highlighted by the black arrows.\label{Figure4}}
\end{figure}

Effective strategies for mode control can be devised by closer inspection of Eq.~(\ref{eq:localgain}), which states that the modal gain depends on three factors:  the gain spectrum, the spatial mode shape, and the dynamic inversion profile. By tuning the overlap between the gain spectrum and the resonance lines of the bright and dark modes one can change their relative intensities. Figure \ref{Figure4}(a) depicts the gain spectrum (shaded area) together with the resonances of the bright and dark modes. The initial choice of $\lambda_{e}=718\,\textrm{nm}$ for the gain maximum implies a strong overlap of the gain spectrum with the bright mode $(713.8\,\textrm{nm})$ and hence a higher gain for this mode. Favored spectrally, the effective gain of the bright mode exceeds that of the dark mode, explaining the lower threshold (Fig.~\ref{Figure3}) and why bright emission sets in first (Fig.~\ref{Figure2}). To examine the impact of spectral alignment, we shift the gain peak relative to the resonances. Figure \ref{Figure4}(b.I) shows the variation of the steady-state output of the bright (circles) and dark modes (squares) depending on the spectral position of the gain peak. We observe a regime of coexistence of the modes between $713\,\textrm{nm}$ and $723\,\textrm{nm}$ (marked by dotted lines). At $723\,\textrm{nm}$, halfway between the resonances, the bright mode switches off. Conversely, below $713\,\textrm{nm}$, with the gain peak shifted towards the lower-wavelength side of the bright resonance, the dark mode switches off and only bright emission occurs. Hence a strong spectral discrimination is required to completely suppress the dark mode owing to its lower losses. These results underpin the necessity of spectral engineering for mode control, which considers both the structural resonances and the fluorescence spectrum. In practice, however, the gain spectrum is often predetermined by the choice of material or application. Instead, one can shift the metamaterial resonances by an appropriate choice of geometric parameters to achieve spectral mode control. 

Spectral tuning is not the only way of controlling the state of emission. The effective gain of a mode is, as Eq.~(\ref{eq:localgain}) indicates, also proportional to the overlap integral of the mode intensity profile and the inversion profile $\int_{V}\,\textrm{d}^{3}\mathbf{r}\left|E(\omega,\textbf{r})\right|^{2}\Delta N(\textbf{r},t)$. While the mode profile is, to a good approximation, determined by the passive metamaterial structure, the inversion profile additionally depends on the pump excitation, in particular on the pump's modal field enhancement that strongly imprints on the inversion. For example, for the specific structure under investigation, the spatial deposition of inversion can be controlled by switching from $E_{x}$- to $E_{y}$-polarized pump light. Figures \ref{Figure4}(c.I) and (c.II) depict the small-signal inversion profiles in the $x$-$y$ plane (before lasing sets in) generated by $E_{x}$- and $E_{y}$-polarized pump light, respectively. When the structure is pumped with $E_{x}$-polarized light, the inversion is distributed in and around the holes with a profile that closely resembles the butterfly shape of the dipolar bright mode [Fig.~\ref{Figure4}(c.I)]. For an $E_{y}$-polarized pump, on the other hand, the inversion concentrates mainly in the projection of the holes as the pump couples to the extraordinary transmission resonance (EOT) at $707.7\,\text{nm}$ for this polarization \citep{Mary2008,Subramina2011,Ramakrishna2011} [Fig.~\ref{Figure4}(c.II)]. The fact that the bright mode benefits less from the inversion deposited by $E_{y}$-polarized pump-light than the dark mode is confirmed by Fig.~\ref{Figure4}(b.II) where we show the steady-state output for $E_{y}$-polarized pumping. A comparison with Fig.~\ref{Figure4}(b.I) reveals that for $E_{y}$-polarized light the dark mode becomes more intense, the bright emission weakens and the window of dark-bright coexistence displays a blueshift.

Finally, we note that the competition between bright and dark modes as well as schemes for mode control depend on the geometry. Using a similar structure with square holes, for example, the $E_{x}$ and $E_{y}$ bright modes are degenerate and the dark mode is then found on the high-frequency side of the bright mode. It is therefore possible to completely suppress the excitation of the dark mode by overlapping its resonance with the absorption line of the gain medium. In this case, dynamic competition between the two bright modes of orthogonal polarization can be observed.\citep{Pusch2012}

In conclusion, we have analyzed the nonlinear dynamic competition of dark and bright lasing states and methods of mode control. Using a full-vectorial Maxwell-Bloch Langevin approach, strong nonlinear mode competition during the transient regime was found, as the modes are harnessing gain in disparate, yet overlapping spatial regions. The steady state input-output curves of the bright and dark modes reveal a characteristic threshold behavior and a regime of stable coexistence. By changing the spectral alignment of the emission line or the spatial deposition of gain, the relative intensities of the modes can be manipulated to the limits where either mode switches off, and sole bright or dark emission occurs. Moreover, in extending the calculations to multiple unit cells, the presented approach will naturally incorporate the competition between Bloch states, opening the possibility to investigate the dynamic build-up of spatial coherence and the emerging radiation patterns. Our work may aid the design of emitting metamaterials by promoting the theoretical understanding of the dynamic interplay between and control of lasing states in active nanoplasmonic structures. 

\begin{acknowledgments}
We gratefully acknowledge financial support provided by the EPSRC, the Royal Academy of Engineering and the Leverhulme Trust.
\end{acknowledgments}



%

\end{document}